\documentclass[12pt,sort&compress]{article}

\usepackage[dvips]{color}
\usepackage{amsmath}
\usepackage{graphicx} 
\usepackage{tabularx}
\usepackage{subfig}
\usepackage{cite}
\usepackage{amsfonts}
\usepackage{tabularx,ragged2e}
\usepackage{mathtools}
\usepackage{bbm}
\usepackage{xcolor}
\newcolumntype{C}{>{\Centering\arraybackslash}X} 
\usepackage[outercaption]{sidecap}    
\usepackage{hyperref}

\begin{document} 
\title{Run-and-tumble particles  on a line with a fertile site}
\date{}
\author{Pascal Grange $^{(\ast)}$,  Xueqi Yao $^{(\ast\ast)}$\\
 Department of Physics, School of Science\\
 Xi'an Jiaotong-Liverpool University\\
111 Ren'ai Rd, 215123 Suzhou, China\\
\normalsize{$^{(\ast)}$ {\ttfamily{pascal.grange@xjtlu.edu.cn}}, $^{(\ast\ast)}$ {\ttfamily{xueqi.yao20@student.xjtlu.edu.cn}}}}
\maketitle

\begin{abstract}
 We propose a model of run-and-tumble particles (RTPs) on a line with a fertile site at the origin. After going through the fertile site, a run-and-tumble particle gives rise to new particles until it flips direction. The process of creation of new particles is modelled by a fertility function (of the distance to the fertile site), multiplied by a fertility rate. If the initial conditions correspond to a single RTP with  even probability
  density, the system is parity-invariant. The equations of motion can be solved in the Laplace domain, in terms of the density of right-movers 
   at the origin. At large time, this density is shown to grow exponentially, at a rate that depends only on the fertility function and fertility rate. Moreover, the total density of RTPs (divided by the density of right-movers at the origin), reaches a stationary state that does not depend 
 on the initial conditions, and presents a local minimum at the fertile site.
\end{abstract}

\tableofcontents

\section{Introduction}

The run-and-tumble particle (RTP) is a simple model of active constituents such as bacteria, including {\emph{E. coli}} \cite{berg2008coli,ramaswamy2010mechanics,cates2015motility}.
  The particle draws energy from its environment  to sustain a motion at constant velocity, in a direction 
   that changes stochastically.  The corresponding equations of motion therefore involve two densities, 
    one for each velocity state. They are coupled, but upon elimination they give rise to the telegrapher's equation. 
     Recent  developments on the RTP in one dimension  include relaxation properties  with
     coupling to diffusion \cite{malakar2018steady}.
     The properties of the random shape of the trajectory of the RTP in dimension two have recently been studied in \cite{hartmann2020convex}.  
     Developments  involving multiple RTPs on a line include \cite{le2019noncrossing}, where exact results on the 
  non-crossing probability of two RTPs have been obtained.
   Models without conservation of the number of particles have been proposed: in  \cite{masoliver1992solutions}  the telegrapher's equation
    was studied in the presence of traps. 
     In  \cite{mori2020universal}  the survival probability of an RTP in presence of an 
      obstacle was worked out  in arbitrary dimension. Moreover, the steady-state probability density of an  RTP subjected to resetting 
       has been obtained in \cite{evans2018run}.
     Exact results using the propagator in  higher dimension have been achieved in \cite{santra2020run,Santra2020} for the RTP subjected to resetting.\\


On the other hand, recent developments \cite{bauer2019random} have given rise to a detailed understanding of the 
 long-time behaviour of free diffusive random walkers on a lattice, whose number is allow to grow through the addition 
  of a fertile site (for earlier results on fertile sites, see \cite{redner1984unimolecular,ben1989random}).
  Random walkers give rise to new random walkers when they are 
  at the fertile site. The random walkers behave like non-interacting diffusive particles. In such a situation the number of particles can only grow. The growth is exponential if the dimension of the lattice is sufficiently 
   low. {{Moreover, dividing the number of random walkers at each site by the total number of random walkers in the system yields a 
    density, which was shown to reach a stationary state.}}\\

{{Due to the proliferating nature of bacteria, collective behaviour of systems of multiple RTPs beg for modelling.  
    Lattice models have been proposed,  providing both insights into collective behaviour and numerical tools to simulate the dynamics of RTPs. For instance,  models of multispecies of random walkers with finite persistence length have been proposed in \cite{thompson2011lattice}, addressing interactions that induce a decrease of the jump rate in response to an increase in density. 
   The size of dense clusters of swimmers on a lattice was shown in \cite{soto2014run} to exhibit scaling properties in terms of the rate of velocity switch. 
    Mutual exclusion of
     two RTPs on a one-dimensional lattice was shown to yield a steady state with a jammed component in \cite{slowman2016jamming}.}}\\

{{In this work we neglect interactions but model the increase of {{density}} in the continuum by allowing the creation of new RTPs.}}
We consider non-interacting run-and-tumble particles on a line with a fertile site.
  A fertile site models a source of nutrient at the origin, that triggers any  passing constituent to give rise to  new constituents (for example by cell division).
In continuous space, modelling a fertile site by the addition of a Dirac mass at the origin (multiplied by the density of particles at the origin)
   gives rise to singularities. Even if the  distribution of particles is absolutely continuous in the initial state on the system,
    it develops a singularity at the fertile site 
    at the origin at positive time, which cannot be multiplied with a Dirac mass in the equations of motion.
     We therefore have to propose a regularisation 
     of the model, replacing the Dirac mass by a smooth fertility function.\\
      
      The paper is organised as follows. In Section 2  we present the model, derive the coupled equations of motion 
       and pick symmetric boundary conditions. In Section 3 we take the  Laplace transform of the equations of motion, which gives rise to a decoupling 
        of left-movers and right-movers. In Section 4 we solve the resulting second-order ordinary differential equation, treating the unknown density of right-movers at the origin as a parameter. The resulting solution yields a constraint on this density of right-movers at the origin:  upon inversion of the Laplace 
         transform, it satisfies an integral equation. In Section 5 this integral equation is used to derive the rate of exponential growth of the density of particles 
          at the fertile site, in a self-consistent way. In Section 6 we normalise the density of RTPs by the density of right-movers at the origin, and 
        work out the large-time limit of this normalised density, which is shown to have a local minimum at the fertile site. In Section 7 we consider limits
         of low and high fertility and illustrate the model 
         for a particular (gamma-distributed) form of the fertility function.


\section{Model and quantities of interest}

 We consider non-interacting run-and-tumble particles on a line (with coordinate at time $\tau$  denoted by  $X(\tau)$),
  whose  velocity  switches stochastically  between $+ v$ and $-v$, for a fixed positive velocity $v$:
 \begin{equation}
  \frac{dX}{d\tau} = v \sigma( \tau ),
 \end{equation}
 where $\sigma$ is a sign that switches according to a Poisson process of intensity $\gamma$. 
   Let us rescale space and time coordinates by choosing $\gamma^{-1}$ as the unit of time and  $\gamma^{-1} v$ as the unit of length:
   \begin{equation}\label{rescaling}
    x:= \frac{X}{\gamma^{-1} v},\;\;\;\;\;\;\;\;\;\;\; t:= \gamma \tau.
   \end{equation}
   With this choice of coordinates, the velocity state of a particle can be $+1$ or $-1$.
Let us denote by $n_\pm(x,t)$ the densities of RTPs with fixed velocity state:
    \begin{equation}
     n_\epsilon(x,t) dx := \{ {\mathrm{average\; number \;of\; RTPs\; at\; time\;}} t\;{\mathrm{ in}}\; [x,x+dx]\;{\mathrm{with\;velocity}}\;\epsilon\},\;\;\;\;{\mathrm{for}}\;\;\epsilon \in \{-1, +1 \}.
    \end{equation}
    We will call $n_+$ (resp. $n_-$) the density of right-movers (resp. left-movers).\\

  Moreover, the origin is a fertile site (as in the model studied in  \cite{bauer2019random}, for diffusive particles on a discrete space): after going through 
     the origin, a constituent can give rise to other constituents. We will assume that particles can pull on a source of nutrient after going through the origin, as if they became hooked to the origin by an elastic 
      band, through which they can pump a nutrient. {{They produce new particles at a rate that depends on the distance they have travelled since going through the origin.}} When they flip  direction after going through  the origin, they stop pulling on the elastic band, and stop creating new particles. They behave as regular RTPs until they go through the origin again.\\

    The creation of RTPs at the fertile site is therefore modelled by adding creation terms to the evolution equation of the 
     equation satisfied by the  density of a single RTP:
     \begin{equation}\label{sys}
     \begin{split}
    \frac{\partial n_+(x,t)}{ \partial t} &=  -\frac{\partial  n_+(x,t)}{\partial x} -   n_+(x,t) + n_- (x,t) + {\textcolor{black}{K e^{-x} \Theta( x )  n_+(0,t-x)}},\\
   \frac{\partial n_-(x,t)}{ \partial t} &=  +\frac{\partial  n_-(x,t)}{\partial x} + n_+(x,t) - n_- (x,t) + {\textcolor{black}{K e^x   \Theta( -x ) n_-(0,t+x)}}.\\
    \end{split}
     \end{equation}
   The function $\Theta$ is a positive function modelling the rate of production of new particles by a particle that has gone through the origin and has not yet changed direction. {{We will call $\Theta$ the fertility function.}} The parameter $K$ is a positive constant. We will call $K$  the fertility rate. The rate of production of particles is conserved if the product $K\Theta$ is conserved.  
     To fix the parameters we can therefore assume that $\Theta$ is normalised:
     \begin{equation}
      \int_0^{\infty} \Theta(x) dx = 1.
     \end{equation}
     Obviously $\Theta(x)=0$ if $x$ is  negative (a constituent cannot start producing new constituents before {{going}} through the fertile site). {{The argument of  the fertility function is $x$ in the creation terms for right-movers (resp. $-x$ for left-movers). In both cases this argument is positive if the creation term is positive: it is the distance travelled from the origin by constituents that have arrived at coordinate $x$ from the origin without flipping direction.}}\\

     To avoid singularities, we will assume that $\Theta$ is smooth. In particular,
    \begin{equation}\label{smoothDer}
     \Theta(0) = \Theta'(0) = 0.
    \end{equation}
   The factor $e^{-|x|}$ inserted in the last term of both equations of motion is the probability that a particle that has gone through the origin at time $t-x$ with positive velocity has not yet switched the sign of its velocity at time $t$ (because particles have unit velocity in our units). {{When a new particle is created, it is introduced into the system at the position of its parent particle.
    Moreover, every new particle is assumed to inherit the velocity of its parent. Hence  $n_+(0,t-x)$
       (resp. $n_-(0,t-x)$) contributes to the time derivative of  $n_+(x,t)$
       (resp. $n_-(x,t)$) in the equations of motion.}} We avoided singularities by not modelling the fertile site by a Dirac mass. 
       If we pick smooth initial conditions, we can therefore assume that the densities of left- and right-movers are smooth functions. 
         We have therefore obtained a non-local modification of the coupled system of equations satisfied by an ordinary RTP. This system 
        is recovered by substituting zero to the fertility rate $K$.\\

   Let us define initial conditions by a smooth and parity-invariant  
  probability density $\varphi$ on the real line:
 \begin{equation}\label{initCond}
  n_+(x,0) = n_-(x,0) = \frac{1}{2}\varphi(x),
 \end{equation}
 where $\varphi$ is a smooth, even probability density on the real line.
  The run-and-tumble particle with  the initial density of left- and right-movers given by a Dirac mass at the origin 
   is well studied (see \cite{othmer1988models,martens2012probability,weiss2002some,evans2018run}), and the corresponding probability density 
    is expressed in terms of Bessel functions, and Dirac masses at the ends of the interval $[-t, t]$ of available positions at time $t$. The Dirac masses keep track of the initial state of the system: they correspond    to  trajectories in which no switching of velocity has taken place since time $0$.
     In  our model we picked a smooth function instead of a Dirac mass
      to define  the initial conditions. 
      This choice ensures that the densities of left-movers and right-movers are 
       absolutely continuous. Moreover, the system is invariant under the parity 
      transformation
      \begin{equation}
       x\mapsto -x,\;\;\;\;\;\;\;\;\; n_\pm \mapsto n_\mp
      \end{equation}
       at all times. Indeed  the initial state of the system is parity invariant, and the equations of motion (Eq. (\ref{sys})) are. 
       We can therefore write
        \begin{equation}\label{parity}
       \forall x,t, \;\;\;\;\; n_-(x,t) = n_+(-x,t),
      \end{equation}
      and  solving the equations of motion in $n_+$ is   enough to provide a solution of the model.\\

 \section{Laplace transform of the equations of motion}
 
 The equations of motion of a single RTP are known to decouple
             upon taking the Laplace transform in the time coordinate (see \cite{evans2018run}). It is therefore natural 
              to apply the same transformation to our model. 
      Let us denote  the Laplace transform of time-dependent quantities as follows:
      \begin{equation}
      \tilde{f}(s) := \int_0^\infty f(t) e^{-st} dt.
      \end{equation}
      The process starts at time zero, so we write  $n_\pm(x,t)=0$ for all $x$ and all negative $t$. 
      The Laplace transform of the creation terms in Eq. (\ref{sys}) reads as follows (for positive $s$):
      \begin{equation}\label{LaplaceBirth}
      \begin{split}
   e^{- x} \Theta(x)  \int_0^\infty  n_+(0,t-x)   e^{-st} dt &=  
      e^{- x}  \Theta( x) \int_{-x}^\infty  n_+(0,u)e^{-s(u+x)}du \\
       &=  e^{-(s+1)x}  \Theta( x) \int_0^\infty  n_+(0,u)e^{-su}du \\
         & =  e^{-(s+1)x} \Theta( x)  \widetilde{n_+}(0,s),\\
     e^{+x} \Theta(-x)    \int_0^\infty  n_-(0,t+x)   e^{-st} dt &=   e^{+ ( s+1) x } \Theta(-x) \widetilde{n_-}(0,s),
      \end{split}
      \end{equation}
      where we used the fact that $n_\pm(0,u) = 0$ for negative time $u$.\\

      The Laplace transform of the equations of motion therefore  reads
     \begin{equation}\label{LaplaceSys}
     \begin{split}
    s\widetilde{n_+}(x,s) -  \frac{1}{2}\varphi(x)&=  -\frac{\partial   \widetilde{n_+}(x,s)}{\partial x} -    \widetilde{n_+}(x,s) +  \widetilde{n_-}(x,s) +  {\textcolor{black}{K\widetilde{n_+}(0,s) \xi(x)}},\\
    s \widetilde{n_-}(x,s)  -   \frac{1}{2}\varphi(x) &=  +\frac{ \partial   \widetilde{n_-}(x,s)}{\partial x} +  \widetilde{n_+}(x,s) -  \widetilde{n_-}(x,s)+
     {\textcolor{black}{ K \widetilde{n_-}(0,s)\xi(-x)}},
    \end{split}
     \end{equation}
     where we used the initial condition  defined in  Eq. (\ref{initCond}) on the l.h.s., and introduced the notation
      \begin{equation}
     \xi(x) :=  e^{-(s+1)x}\Theta(x).
       \end{equation}
      Taking the derivative w.r.t. $x$ of Eqs (\ref{LaplaceSys}) and rearranging 
      yields
       
      \begin{equation}\label{secDer}
    \partial_x^2   \widetilde{n_+}(x,s)  + (s+1)\partial_x\widetilde{n_+}(x,s) - \partial_x \widetilde{n_-}(x,s) =   \frac{1}{2}\varphi'(x) + {\textcolor{black}{ K\widetilde{n_+}(0,s)  \xi'(x)}}.
     \end{equation}
     
%
%
%
%
%
     
    Using the Laplace transform of the  equations of motion (Eq. (\ref{LaplaceSys})) we obtain
      \begin{equation}
      \begin{split}
      (s+1)\partial_x\widetilde{n_+}(x,s) - \partial_x \widetilde{n_-}(x,s)=& 
        (s+1)\left[    - s\widetilde{n_+}(x,s) +  \frac{1}{2}\varphi(x)    -    \widetilde{n_+}(x,s) +  \widetilde{n_-}(x,s) + {\textcolor{black}{ K \widetilde{n_+}(0,s) \xi(x)    }}\right]\\
      &- s \widetilde{n_-}(x,s)  +  \frac{1}{2}\varphi(x)+  \widetilde{n_+}(x,s) -  \widetilde{n_-}(x,s)+
       {\textcolor{black}{ K\widetilde{n_-}(0,s)\xi(-x)}}\\
       =& (s+1)\left[    - s\widetilde{n_+}(x,s) +  \frac{1}{2}\varphi(x)    -    \widetilde{n_+}(x,s)  +{\textcolor{black}{  K \widetilde{n_+}(0,s) \xi(x) }}   \right]\\
      &+  \frac{1}{2}\varphi(x)+  \widetilde{n_+}(x,s)+
     {\textcolor{black}{ K  \widetilde{n_-}(0,s)\xi(-x)}}\\
       =& -s(s+2) \widetilde{n_+}(x,s) + \left( \frac{s}{2} + 1\right) \varphi(x)\\
         &+  {\textcolor{black}{ K\widetilde{n_+}(0,s) (s+1) \xi( x )}}
        +   {\textcolor{black}{ K\widetilde{n_-}(0,s)\xi(-x)}}.
     \end{split}
         \end{equation}
       Substituting into Eq. (\ref{secDer})  yields
            \begin{equation}\label{aboveDiff}
            \begin{split}
    \partial_x^2   \widetilde{n_+}(x,s)  - s(s+2) \widetilde{n_+}(x,s) =& - \left( \frac{s}{2} + 1\right) \varphi(x) + \frac{1}{2}\varphi'(x)\\
         &- {\textcolor{black}{ K \widetilde{n_+}(0,s) (s+1) \xi( x )}}
        - {\textcolor{black}{ K \widetilde{n_-}(0,s)\xi(-x)  + K\widetilde{n_+}(0,s)  \xi'(x)}},
         \end{split}
     \end{equation}
     which almost displays the expected decoupling, except for the  Laplace transform of the density of left-movers at the origin $\widetilde{n_-}(0,s)$,
      which appears on the r.h.s, and can be re-expressed using the parity symmetry of the model.
       Indeed,  Eq. (\ref{parity}) holds at the fertile site $x=0$. Let us denote the common value of the densities of left- and right-movers at the origin and 
        at time $t$ by $R(t)$:
   \begin{equation}
    R( t) :=  n_+(0,t) =  n_-(0,t). 
   \end{equation}
         We can therefore rewrite Eq. (\ref{aboveDiff}) as follows:
           \begin{equation}\label{diff2}
    \partial_x^2   \widetilde{n_+}(x,s)  - s(s+2) \widetilde{n_+}(x,s) = g_+(x,s),
        \end{equation}
        where the function $g_+$ is an affine function of the unknown density of right-movers at the fertile site, with coefficients 
         expressed in terms of the initial conditions and the other  parameters of the model (fertility function $\Theta$ and fertility rate $K$):
          \begin{equation}\label{defg}
          \begin{split}
    g_+(x,s) :=& 
     - \left( \frac{s}{2} + 1\right) \varphi(x) + \frac{1}{2}\varphi'(x)
         +{\textcolor{black}{ K\tilde{R}(s) \left[ -   (s+1)  \xi( x )
        -  \xi(-x)  +   \xi'(x) \right]}}\\
        =&    - \left( \frac{s}{2} + 1\right) \varphi(x) + \frac{1}{2}\varphi'(x)\\
        &
         +{\textcolor{black}{ K  \tilde{R}(s) \left[ -   2(s+1)  e^{-(s+1)x} \Theta(x)
        -   e^{(s+1)x} \Theta(- x)  + e^{-(s+1)x} \Theta'(x) \right]}}.
        \end{split}
     \end{equation}
          The function $g_+$ is a  smooth function of $x$ because the fertility rate $\Theta$ is.
            We can attempt to solve this equation as a 
             second-order ordinary differential equation, treating $\tilde{R}(s)$ as a parameter. The solution will yield a 
              consistency condition satisfied by the density of right-movers at the origin.

\section{Integration of the equations of motion}

  If we treat   the Laplace variable $s$ conjugate to time as a constant parameter, Eq. (\ref{diff2}) becomes a second-order ordinary differential equation 
   of the form
\begin{equation}
y''(x) - \sigma y(x) = f(x),
 \end{equation}
 with the notations
 \begin{equation}
 \begin{split}
 \sigma &:= s(s+ 2),\\
 f(x) &:= g_+(x,s),
 \end{split}
 \end{equation}
 {{where the function $g_+$ is defined in Eq. (\ref{defg}) in terms of the parameters of the problems (initial conditions, fertility rate  and fertility function),   and of the unknown density of right-movers at the origin in Laplace space, denoted by $\widetilde{R}(s)$}}.\\

  This differential  equation is readily reformulated as a first-order  equation
  in the vector $Y(x)$ defined as  
\begin{equation}
Y(x) := 
\begin{bmatrix} 
  y(x)\\
 y'(x)
\end{bmatrix}. 
 \end{equation} 
{{The problem reads}}
\begin{equation}\label{ODE1}
Y'(x) = M Y(x) + F(x),
 \end{equation}
with
\begin{equation}
M := 
\begin{bmatrix} 
  0  & 1 \\
   \sigma  & 0 \\
\end{bmatrix},\;\;\;\;\;\; 
F(x) := \begin{bmatrix} 
  0   \\
   f(x) 
\end{bmatrix}.
 \end{equation}
{{ This problem is readily solved by diagonalising the matrix $M$. The derivation is shown in Appendix A.
The  density of right-movers is expressed as
 \begin{equation}\label{affineSys}
 \widetilde{n_+}(x,s) =  \cosh( x\sqrt{\sigma} ) \lambda (s) +  \frac{1}{\sqrt{\sigma}} \sinh( x\sqrt{\sigma} ) \mu(s) + \frac{1}{\sqrt{\sigma}} \int_{0}^x \sinh\left( (x-y)\sqrt{\sigma}\right) g_+(y) dy,
\end{equation}
  where $\lambda(s)$ and $\mu(s)$ are integration constants, expressed in Eqs (\ref{lambdamuDef}) in terms of the function $g_+$.}}
 The {{resulting}} density of right-movers (in Laplace domain) is  an affine function of $\widetilde{R}(s)$,  
  because this  unknown quantity enters the definition of the function $g_+$ in Eq. (\ref{defg}), with coefficients that depend on the parameters 
   of the fertile site (and not on the initial conditions).
   We can therefore  rewrite  Eq. (\ref{affineSys}) at $x=0$ in the following form:
   \begin{equation}\label{LaplacePack}
   \widetilde{n_+}(0,s) = \widetilde{R}(s)= \lambda(s) =   \widetilde{\psi}(s) + \widetilde{\Xi}( s) \widetilde{R}(s),
   \end{equation}
    where the $s$-dependent coefficients  $\widetilde{\psi}(s)$ and $ \widetilde{\Xi}( s)$ have been denoted as Laplace transforms.
     Let us extract the following expressions from the value of $\lambda(s)$ obtained in the Appendix (Eq. (\ref{lambdaVal})):
  \begin{equation}\label{psiTilde}
    \widetilde{\psi}(s) = {{\frac{1}{2}\sqrt{\frac{s+2}{s}} \int_0^\infty e^{-y\sqrt{s(s+2)}}\varphi(y) dy}},
   \end{equation} 

   \begin{equation}\label{exprXi}
   \widetilde{\Xi}( s) = \frac{ K  }{2\sqrt{s(s+2)}}\left(  s + 2 - \sqrt{s(s+2)} \right) \tilde{\Theta}(s+ 1 +\sqrt{s(s+2)} ).
   \end{equation}

%
  
%
%

    We have therefore obtained a formal solution of the problem in the Laplace domain, in terms of the unknown density of particles at the 
     origin. 
  Inverting the Laplace transform maps the ordinary product  to a convolution product. The affine dependence of the r.h.s.  of Eq. (\ref{LaplacePack})  
   on the density of left- and right-movers at the origin therefore yields a consistency condition on the density 
      in the form of an integral equation:
  \begin{equation}\label{consistency}
    R(t) =   \psi(t) +  \int_0^t \Xi( t-u) R(u) du.
   \end{equation}
   
  \section{Exponential growth of the number of particles}
   
   {{In a zero-dimensional model of growth, non-interacting particles sit {{ on top of  a fertile site}}. Each particle produces offspring at a constant rate,
    corresponding to the amount of energy it can extract from the fertile site. If we model this amount as a constant, the number of particles grows exponentially.
     In our one-dimensional model, the production of particles can happen anywhere on the real line, but it is more likely to occur close to the fertile site, because 
      particles stop producing offspring when they flip direction after going through the fertile site. The growth of the number of particles is therefore expected to be governed by the behaviour of the model close to the fertile site, which motivates us to expect an exponential growth, as in the zero-dimensional model. Moreover,  in dimension one, a population of  diffusive random walkers was shown in \cite{bauer2019random} to grow exponentially.}}\\

       Let us {{therefore}} look in a self-consistent way for an exponential equivalent of the density of right movers at large time. 
   We {{postulate the existence of two  positive 
  constants}} $\rho$ and $\chi$ (independent of both $x$ and $t$), such that
  \begin{equation}\label{Ansatz}
   R(t) \underset{t \to \infty}{\sim} \rho e^{\chi t}.
    \end{equation}

 {{  We have to take the large-time limit of the 
  consistency condition (Eq. (\ref{consistency})) satisfied by the density of right-movers at the origin.
     The function $\psi(t)$ is bounded  because it is the density of right-movers at the origin if the fertility rate $K$ is set to zero (in which case there is only one particle in the system).  At large time, the r.h.s. of Eq.  (\ref{consistency}) is therefore equivalent to the integral term:
       \begin{equation}
    R(t) \underset{t \to \infty}{\sim}   \int_0^t \Xi( t-u) R(u) du.
   \end{equation}
   Let us inject the  exponential growth postulated in Eq. (\ref{Ansatz})  on both sides, and rescale the  integration variable by introducing  $v:= t^{-1}u$:
\begin{equation}
   e^{\chi t} \underset{t \to \infty}{\sim} t  \int_0^1 \Xi( t(1-v) ) e^{\chi t v} dv.
    \end{equation} 
 In the large-time limit, dividing both sides of the above equivalent by $e^{\chi t}$ yields the limit
  \begin{equation}
   \underset{t\to\infty}{\lim} t  \int_0^1 \Xi( t(1-v) ) e^{\chi t (v-1)} dv = 1.
  \end{equation}
  Changing the integration variable to $T := t(1-v)$, the l.h.s. of the above equation becomes the Laplace transform of the function $\Xi$, taken at the 
   unknown rate $\chi$:
    \begin{equation}
   \underset{t\to\infty}{\lim}  \int_0^t \Xi( T) e^{-\chi  T} dT= 1.
  \end{equation}
  The rate $\chi$  therefore satisfies 
   \begin{equation}\label{defChi}
   \tilde{\Xi}( \chi)= 1.
  \end{equation}
   Using the expression of the Laplace transform $\widetilde{\Xi}$ in Eq. (\ref{exprXi}), we obtain 
  an equation in $\chi$, the postulated rate of exponential growth:
  \begin{equation}\label{eqRate}
  1=    \frac{K}{2}  \left(  \sqrt{1+\frac{2}{\chi}} - 1 \right) \tilde{\Theta}(\chi+ 1 +\sqrt{\chi(\chi+2)} ),
   \end{equation} 
    As the fertility function $\Theta$ is positive, the Laplace transform $\tilde{\Theta}(s+ 1 +\sqrt{s(s+2)} )$
     is a positive and decreasing function of $s$. The quantity $\tilde{\Xi}( s)$ is therefore a decreasing function of $s$.
  Moreover, we have the following two asymptotic behaviours:
   \begin{equation}
   \tilde{\Xi}(s)\underset{s\to 0^+}{\sim} \frac{K\tilde{\Theta}( 1)}{s\sqrt{2}}\underset{s\to 0^+}{\longrightarrow}+\infty, \;\;\;\;\;\;\;  \tilde{\Xi}(s)\underset{s\to +\infty}{\longrightarrow} 0.
   \end{equation}
   There is therefore a unique positive solution to  Eq. (\ref{defChi}), which depends on the choice of parameters $K$ and $\Theta$, but not on the initial conditions.\\
}}

  {{  
  On the other hand, the Laplace transform $\tilde{R}(s)$  of right-movers at the origin is obtained from  Eq. (\ref{LaplacePack}) as
  \begin{equation}
   \tilde{R}(s) = \frac{\tilde{\psi}(s)}{1- \tilde{\Xi}(s)}.
 \end{equation}
 Because of Eq. (\ref{defChi}), this expression has a pole at $s=\chi$. {{The Taylor expansions of $\tilde{\psi}$ and $\tilde{\Xi}$ about the point $\chi$  read}}
 \begin{equation}\label{comp1}
 \begin{split}
 \tilde{\Xi}( \chi + h) &= 1 + h( \tilde{\Xi})'(\chi)+ o(h),\\
 \tilde{\psi}(\chi + h) &= \tilde{\psi}(\chi)( 1 + o(1)).
 \end{split}
 \end{equation}
 {{An equivalent of $\tilde{R}( \chi + h)$ when $h$ goes to zero is therefore obtained as}} 
  \begin{equation}\label{comp1}
 \begin{split}
   \tilde{R}(\chi + h) &= -\frac{\tilde{\psi}(\chi)}{h\tilde{\Xi}'(\chi) }( 1 + o(1)).
   \end{split}
 \end{equation}

   The asymptotic behaviour of the density of right-movers at the origin described in Eq. (\ref{Ansatz})
   can be expressed as
  \begin{equation}
   \underset{t\to \infty}\lim U(t) = 1,\;\;\;\;\;\;\;{\mathrm{with}}\;\;\;\;\;\;\;\;U(t):= \rho^{-1} e^{-\chi t}R(t).
  \end{equation}
  Applying the final-value theorem to the function $U$ yields
   \begin{equation}\label{comp2}
   \tilde{U}(h) \underset{h\to 0^+}{\sim} \frac{1}{h}.
  \end{equation}
  On the other hand, the Laplace transform of the function $U$ is readily expressed in terms of the Laplace transform  of $R$:
  \begin{equation}\label{comp3}
    \tilde{U}(h) = \rho^{-1}\int_0^\infty e^{-(\chi +h)t}R(t) dt =  \rho^{-1} \tilde{R}( \chi +h).
  \end{equation}
   Consistency between Eqs (\ref{comp1},\ref{comp2},\ref{comp3}) yields
    \begin{equation}
    \rho = -\frac{\tilde{\psi}(\chi)}{\tilde{\Xi}'(\chi) }.
   \end{equation}
   The prefactor $\rho$ is positive because $\tilde{\Xi}$  is a decreasing function, and $\tilde{\psi}$ is positive (which is manifest from Eq. (\ref{psiTilde}) because $\varphi$ is non-negative as a probability density). Moreover, $\rho$ depends on the initial condition $\varphi$ through the numerator in the above expression.
}}

 \section{Large-time behaviour of the spatial distribution of\\ right-movers}
  Let us come back to Eq. (\ref{affineSys}) satisfied by the density of right-movers $\widetilde{n_+}(x,s)$. It is an affine function of the Laplace transform 
   of the density of right-movers at the origin (denoted by $\tilde{R}(s)$), but the coefficients depend on both the coordinate $x$ 
    and the Laplace variable $s$. We can therefore write
    \begin{equation}\label{toReadOff}
     \widetilde{n_+}(x,s) = \tilde{\nu}(x,s) +  \tilde{M}(x,s)\tilde{R}(s),
    \end{equation}
    where the coefficients $\tilde{\nu}$  and $\tilde{M}$ have been denoted as Laplace transforms.
    In Eq. (\ref{LaplacePack}) we used a special version of this equation for $x=0$, with coefficients given by the special values $\tilde{\psi}(s) = \tilde{\nu}(0,s)$
     and $\tilde{\Xi}(s) = \tilde{M}(0,s)$. 
     Inverting the Laplace transform maps the ordinary product to a convolution in time:
     \begin{equation}\label{n+xt}
      n_+(x,t) =  \nu(x,t) + \int_0^t   M(x,t-l) R(l) dl.
     \end{equation}
      The first term  $\nu(x,t)$ is the value of the density $n_+(x,t)$ if  the fertility rate $K$ is set to zero (in this case the number of particles 
       is conserved, and the density is bounded).
        If $K$ is non-zero, the second term in Eq. (\ref{n+xt}) dominates at large time because of the 
        exponential growth of the density $R$.
     If we  compare  the density of right-movers at position $x$ to the density at the origin, and take the large-time limit, 
      we therefore obtain the following equivalent (upon the change of variable defined by $z:= t-l$):
       \begin{equation}
     \frac{  n_+(x,t)}{R(t)}  \underset{t\to \infty}{\sim}\frac{1}{R(t)}\int_0^t   M(x,z) R(t-z) dz \underset{t\to \infty}{\sim}\int_0^t   M(x,z) e^{-\chi z} dz
     =   \tilde{M}(x,\chi).
     \end{equation}
     The  ratio of the density of right-movers at $x$ to the density of particles at the origin therefore reaches a stationary state.
       We can read it off by extracting the coefficient 
        of $\tilde{R}(s)$ from the following equation (obtained by substituting the growth rate $\chi$ to the Laplace variable $s$  in   Eq. (\ref{toReadOff})):
    \begin{equation}\label{spatial}
    \begin{split}
 \widetilde{n_+}(x,\chi) = & \cosh( x\sqrt{\chi(\chi+2)} ) \lambda (\chi) +  \frac{1}{\sqrt{\chi(\chi + 2)}} \sinh( x\sqrt{\chi(\chi + 2 )} ) \mu(\chi) \\
  &+ \frac{1}{\sqrt{\chi(\chi+2)}} \int_{0}^x \sinh\left( (x-y)\sqrt{\chi(\chi + 2)}\right) g_+(y,\chi) dy.
 \end{split}
\end{equation}
 We know from the definition of $\lambda(s)$ and $\chi$ in Eqs (\ref{lambdamuDef}) and (\ref{defChi}), and from the decomposition of $\lambda(s)$ in
  Eq. (\ref{LaplacePack}) that the coefficient of $\tilde{R}(s)$ contributed by $\lambda(\chi)$ in Eq. (\ref{spatial}) equals $1$. We need to extract the coefficient 
    of ${\textcolor{black}{\tilde{R}(\chi)}}$  in the integral term in Eq. (\ref{spatial}). With the notations
    \begin{equation}\label{notationsJ}
    \begin{split}
    \mathcal{I}(x,s) :=&  \int_{0}^x \sinh\left( (x-y)\sqrt{\sigma}\right) g_+(y) dy\\
   =& k(s) + J(\chi,x) \tilde{R}(s),
 \end{split} 
    \end{equation}
   we express $\tilde{M}$ as 
   \begin{equation}\label{stationaryx}
         \begin{split}
      \tilde{M}(x,\chi) = & \cosh\left(  x \sqrt{\chi(\chi + 2 )} \right)  +  \frac{1}{\sqrt{\chi(\chi + 2)}} \sinh( x \sqrt{\chi(\chi + 2 )}) \xi(\chi) 
      +\frac{1}{\sqrt{\chi( \chi + 2)}} J(\chi,x),\\
      \end{split}
    \end{equation}
      where $\xi( \chi)$ is the coefficient of ${\textcolor{black}{\tilde{R}(\chi)}}$ in $\mu( \chi)$   (worked out in Eq. (\ref{muApp}) ), and $J(\chi,x)$ is 
       worked out in Eq. (\ref{JApp}). Substituting these expressions into Eq. (\ref{stationaryx}) yields
     \begin{equation}\label{stationaryxz}
         \begin{split}
      \tilde{M}(x,\chi) = & \cosh\left(  x \sqrt{\chi(\chi + 2 )} \right)  \\
     & +  \frac{K}{2\sqrt{\chi(\chi + 2)}} \sinh( x \sqrt{\chi(\chi + 2 )})  (\chi- \sqrt{\chi(\chi+2)}) \tilde{\Theta}(\chi+ 1 +\sqrt{\chi(\chi+2)} )  \\
      &+\frac{K}{2\sqrt{\chi( \chi + 2)}}   \theta( x) [ 
      ( -\chi -1+\sqrt{\chi (\chi +2)}) e^{x\sqrt{\chi( \chi + 2 )}}\int_0^x   e^{-{y( \chi  + 1 +\sqrt{\chi (\chi +2)})}} \Theta(y) dy
   \\
    & + ( \chi +1+\sqrt{\chi (\chi +2)})  e^{-x\sqrt{\chi( \chi + 2 )}} \int_0^x   e^{-y( \chi  + 1 -\sqrt{\chi (\chi +2)})} \Theta(y) dy
         ]\\
     & +\frac{K}{2\sqrt{\chi( \chi + 2)}}   \theta( -x)  [  e^{x\sqrt{\chi( \chi + 2 )}}
       \int_0^{-x}   e^{ -y(\chi+1 -\sqrt{\chi (\chi +2)})}
    \Theta( y)     dy   \\
  &   - e^{-x\sqrt{\chi( \chi + 2 )}}
      \int_0^{-x}  e^{-y( \chi  + 1 +\sqrt{\chi (\chi +2)})} \Theta(y)     dy ],\\
      \end{split}
    \end{equation}
    where $\theta$ denotes the Heaviside step function.\\

%
%
%
%
%
%


\section{Total density  of particles}  
  
   Let us consider the total density of particles, normalised by the density of right-movers at the origin.
    It reaches a steady state denoted by $\mathcal{N}$
     \begin{equation}
      \mathcal{N}(x) := \underset{t\to \infty}{\lim} \frac{n_+(x,t) + n_-(-x,t)}{R(t)} = \tilde{M}(x,\chi) + \tilde{M}(-x,\chi). 
     \end{equation}

 By construction $\mathcal{N}$ goes to zero at both infinities. Moreover, it has an extremum at the origin:
  \begin{equation}
      \mathcal{N}'(0) = \frac{\partial\tilde{M}}{\partial x}(0,\chi) - \frac{\partial\tilde{M}}{\partial x}(0,\chi) = 0. 
     \end{equation}
 
 The nature of this extremum depends on the sign of the second derivative of the stationary density of right-movers 
  at the origin:
   \begin{equation}
      \mathcal{N}''(0) = 2 \frac{\partial^2\tilde{M}}{\partial x^2}(0,\chi). 
     \end{equation}
     
 To work out the above derivative, we need a Taylor expansion of $\tilde{M}(x,\chi)$ around the origin. 
 The derivatives of the step function in the expression  of $\tilde{M}(x,\chi)$ (Eq (\ref{stationaryxz})) do not give rise to singularities, because the corresponding Dirac masses 
   are 
  weighted by coefficients of the form $C_\alpha(0)$ and $C'_\alpha(0)$,
  with the notation
  \begin{equation}\label{prop1}
   C_\alpha(x) = \int_{0}^x \Theta( y ) e^{-\alpha y} dy,  \;\;\;\;\;\;C_\alpha(0) = 0,\;\;\;\;\; C'_\alpha(0) = \Theta(0) = 0,
  \end{equation}
   where $\alpha$ takes the values $\chi + 1 \pm\sqrt{\chi(\chi+2)}$.
  On the other hand,
  \begin{equation}\label{prop2}
  C''_\alpha(0) = -\alpha \Theta(0 ) + \Theta'(0) = 0, 
  \end{equation}
   because the fertility function $\Theta$ is assumed to have a continuous first derivative (Eq. \ref{smoothDer}).
  Hence we can Taylor expand around the origin the expression of $\tilde{M}(x,\chi)$  obtained in Eq. (\ref{stationaryxz}) and read off
  \begin{equation}
    \mathcal{N}''(0) = 2\chi(\chi+2) >0,
  \end{equation}
   as all  the terms of order $x^2$ come from the expansion of $\cosh(x\sqrt{\chi(\chi + 2)})$ (because of the local  properties of the integral terms displayed
    in Eqs (\ref{prop1},\ref{prop2})).
      The steady density profile  of the total number of particles therefore presents a minimum at the origin. The minimum is sharper 
   when  the growth rate of the number of particles is larger. Moreover, the above results remain unchanged if we just assume the fertility 
    function to have a continuous first derivative (without necessarily being {{infinitely}} differentiable).

\subsection*{Example: Gamma-distributed fertility function}

The Gamma density 
\begin{equation}\label{GammaDef}
\Theta(x):= \frac{1}{\Gamma( k) a^k} x^{k-1} e^{-\frac{x}{a}} \mathbbm{1}(x\geq 0 )
\end{equation}
  has a continuous first derivative if $k>2$. 
  {{The Gamma density therefore satisfies the assumptions we made on the fertility function. Moreover, it appeared in models of protein concentration in live cells \cite{friedman2006linking}. It represents  the steady 
  state of the probability distribution $p(c)$ of the concentration $c$ of a given protein molecule in a population of cells. Processes such as cell growth and cell division decrease the concentration, and production of molecules increases the concentration. The production is modelled by random bursts (corresponding to the life cycle of an mRNA molecule that gets translated until its degradation). If the burst does not depend on the number of protein molecules initially present, the production term takes the form of  an integral in the  steady-state equation:
  \begin{equation}\label{geneExp}
  \frac{\partial}{\partial c}[ c p(c) ] + k \int_0^c \left(   \nu(c-c') - \delta( c-c') \right) dc' = 0.
  \end{equation}
   The coefficient $k$ models cell-cyle processes (such as growth, division and degradation of protein molecules), the $\delta$-function accounts for the loss of probability at concentration, and the kernel denoted by $\nu$ 
    describe the distribution of burst sizes. In  \cite{friedman2006linking}, burst sizes were modelled by an exponential distribution
    \begin{equation}\label{expa}
     \nu( y ) := \frac{1}{a} \exp\left( -\frac{y}{a}\right),
    \end{equation}
    motivated by measurements of burst sizes in bacterial cells \cite{cai2006stochastic,yu2006probing}. The convolution kerel in Eq. (\ref{geneExp}) is therefore known, which allows for a solution in Laplace space. Transforming back to ordinary space yields $p(c) = \Theta(c)$, with the expression displayed 
    in Eq. (\ref{GammaDef}).}}\\
    
     {{
    In our model, the coordinate $x$ is not a concentration but a spatial coordinate. However, we may 
     construct an analogy with the  model of  \cite{friedman2006linking} in terms of a bath of active particles whose effect on RTPs  would give rise to  Eq. (\ref{GammaDef}).
       Consider another class of non-interacting active particles  (not the RTPs)
      on the half-line $x>0$, which are drawn to the fertile site by a force $-\kappa x$ (with positive $\kappa$),   and subjected to a friction
       force $-\zeta \dot{x}$ (with positive $\zeta$). Let there be a large and constant number of these particles in the system.
        Without a random element, in the case of a strong friction force,  this would yield a current of particles towards the fertile site, corresponding to particles   with a velocity  $-\zeta^{-1}\kappa x$. Let us  allow these particles to jump towards higher values of $x$, by random amounts distributed 
        exponentially according to Eq. (\ref{expa}), where $a$ is now a fixed length scale, as in Eq. (\ref{GammaDef}). At steady state, the resulting current towards 
         higher values of $x$ 
     compensates the current towards the fertile site.  The density of these additional particles will satisfy Eq. (\ref{geneExp}) on the positive half-line (with $k:=\zeta \kappa^{-1}$).  It will therefore be given by a Gamma density, where $k$ summarises the properties of the current towards the fertile site.\\}}
     
      {{
     This mapping provides a microscopic realisation of the process (described in Section 2) in which RTPs {\emph{pull on an elastic band after   going through the fertile site}}, and produce new particles
      at a rate $\Theta$ set by own internal stochastic process. Instead of relying on this description, we can imagine that the system contains a second class of particles (``activators''). These activators 
       are attracted to the fertile site and subjected to a strong friction force, and may {{propel}} themselves further from the fertile site, with exponentially-distributed  jumps. An activator can interact  with an RTP whose velocity has a sign opposite to its own and has not switched velocity since going through the fertile site. When it meets such an RTP, it triggers the birth of a new RTP. The fertility rate $K$ describes the concentration of activators in the system.}}

  To obtain the steady-state density profile we need to evaluate the following integral (for positive $s$ and $X$):
  \begin{equation}
  \begin{split}
   C_s(X) &= \int_0^X e^{-sy}\Theta(y) dy =  \frac{1}{\Gamma( k) a^k} \int_0^X y^{k-1} e^{-y\left(s + a^{-1}   \right) }dy\\
     &= \frac{1}{\Gamma( k) a^k\left(s + a^{-1}   \right)^k}\int_0^{X(s+a^{-1})} z^{k-1} e^{-z }dz\\
     &=  \frac{1}{\Gamma( k) \left(sa + 1   \right)^k} \gamma( k, X(s + a^{-1})),
   \end{split}
  \end{equation}
 where we denoted the incomplete Gamma function by 
 \begin{equation}
  \gamma(k,Y)= \int_0^Y x^{k-1}e^{-x}dx.
 \end{equation}

 \begin{figure}\label{figureTotalDensity}
\includegraphics[width=18cm]{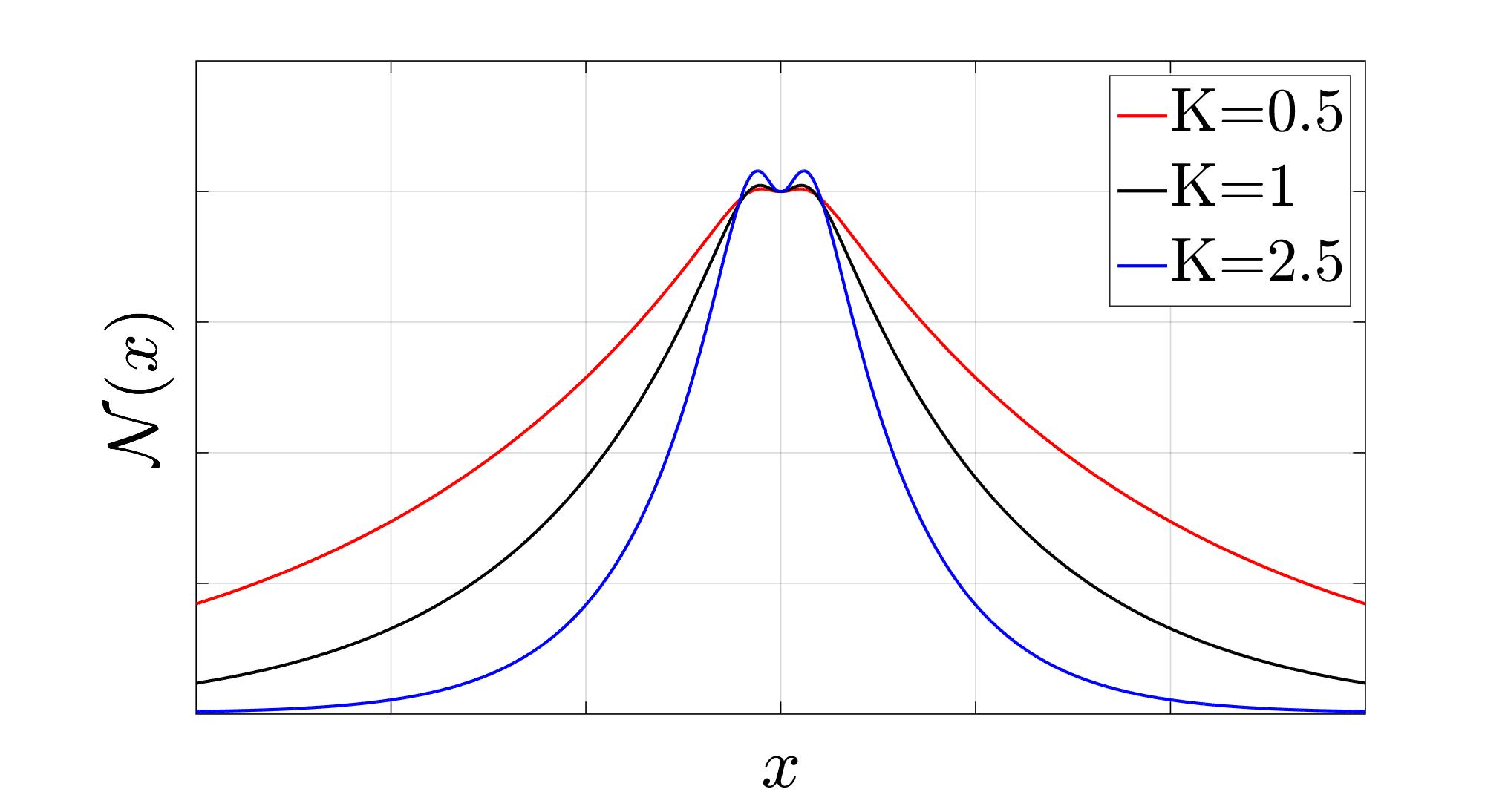} 
\caption{The steady-state profile of the total density of particles, normalised by the density of RTPs, normalised by the density of right-movers at the origin. The fertility function is a Gamma density with $a=1$ and $k=3$. Larger fertility rates induces a faster growth of the total number of particles, which results in a sharper minimum at the fertile site, and in a faster decay at infinity.}
\label{figureTotalDensity}
\end{figure}
 
  To estimate the rate of growth of the number of particles we need the Laplace transform of the Gamma density:
  \begin{equation}\label{LaplaceGamma}
  \tilde{\Theta}(s) = \frac{1}{\left(sa + 1   \right)^k}.
  \end{equation}
 The growth rate $\chi$ is therefore given by the solution of 
 \begin{equation}
  1 = \frac{K}{2}\left(  \sqrt{1+\frac{2}{\chi}} -1 \right) \frac{1}{\left(  [\chi + 1 + \sqrt{\chi(\chi + 2)}]a + 1 \right)^k}.
 \end{equation}
  {{The normalised density profile of RTPs decays exponentially at both infinities:
 \begin{equation}
  \tilde{M}(x,\chi) \underset{|x|\to +\infty}{\sim} e^{-\sqrt{\chi(\chi + 2 )}|x|},
 \end{equation}
  which is  checked in Appendix C. For numerical illustration we picked $a=1$ (which adjusts the scale of the Gamma distribution to the 
   average length travelled by an RTP from the fertile site without switching velocity).  The total density profile is illustrated  in Fig. (\ref{figureTotalDensity}).\\
 }}\\

\subsection*{Low fertility rate}
 If  the fertility rate $K$ is close to zero, the rate $\chi$ is close to zero, which is intuitive, and necessary 
   for both sides of Eq. (\ref{eqRate}) to remain constant in the limit $K\ll 1$:
  \begin{equation}
   1 = \frac{K}{\sqrt{2\chi}}\tilde{\Theta}(1) + o(1).
  \end{equation}
  Hence the growth rate reads
  \begin{equation}
  \chi  \underset{K\to 0}{\simeq} \frac{(K \tilde{\Theta}(1))^2}{2} = \frac{K^2}{2(a+1)^{2k}}.
  \end{equation}
  For low values of the fertility rate, the rate of exponential growth of the density of particles at the origin is therefore quadratic in the 
   fertility rate. This quadratic  behaviour does not depend on the choice of the fertility function 
   (only the coefficient $\tilde{\Theta}(1)^2$ does). Moreover, the second derivative of the stationary density profile of the total number of particles also goes to zero at low fertility rate: 
   \begin{equation}
  \mathcal{N}''(0)\underset{K\to 0}{\simeq}4 \chi \simeq  2 K^2 \tilde{\Theta}(1)^2 = \frac{2K^2}{(a+1)^{2k}}.
  \end{equation}
   The power-law behaviour in the fertility rate is again independent of the choice of fertility fuction in the model.

\subsection*{High fertility rate}
If the fertility rate is high (for a fixed fertility function), the growth rate $\chi$  becomes large,
 using the expression of the Laplace transform of the Gamma density in Eq. (\ref{LaplaceGamma}) yields
\begin{equation}
 \tilde{\Theta}( \chi + 1 + \sqrt{\chi( \chi + 2)}) \underset{K\gg 1 }{\simeq} \frac{1}{(2a\chi)^k},
 \end{equation}
 and Eq. \ref{eqRate} yields
 \begin{equation}
  1 \simeq  \frac{K}{2\chi(2a\chi)^{k}},\,\,\,\,\,\,\,{\mathrm{i.e.}}\,\,\,\,\,\,\,\,\,\chi \underset{K \to \infty}{\sim} \frac{1}{2 a^{\frac{k}{k + 1}}}  K^{\frac{1}{k + 1}},
 \end{equation}
  so that the second derivative of the normalised density profile at the origin becomes large at large fertility, as
  \begin{equation}
  \mathcal{N}''(0)\underset{K\gg a^{-1}}{\simeq}\frac{K^{\frac{1}{k+1}}}{2 a^{\frac{k}{k+1}}}.
  \end{equation}

\section{Discussion}

In this work we have proposed a model of a run-and-tumble particle with a fertile site.  Singularities 
 were avoided by considering smooth initial conditions and a sufficiently regular fertility function (on the other hand, solutions of the equations of motion for a single  RTP  worked out in  \cite{malakar2018steady,evans2018run,santra2020run,Santra2020} assume that the initial configuration is a Dirac mass). The model contains three parameters: the fertility rate $K$, the fertility function $\Theta$ (a  normalised density with continuous first derivative, whose support 
   is on the positive part of the real line), and the initial value $\varphi$ of the density of particles (a smooth, even  probability density on the real line). 
  Moreover, the symmetry of the initial  conditions induces parity symmetry and allows to solve the equations
   of motions for right-movers. The model is considerably simplified by assuming that a particle loses the ability to emit new particles after changing 
    direction. We obtained the rate of exponential growth of the density of right-movers at the origin as the unique solution 
     of an equation involving the fertility rate and the fertility function. This rate of growth is therefore independent of the initial probability density 
      $\varphi$. On the other hand, the equivalent of the density of righ-movers at large times contains a  prefactor whose value does depend on the initial conditions.  \\

 The fertility rate is zero at the 
   fertile site, which is a consequence of the smoothness of the fertility function. In the example of the gamma distribution, 
    the fertility function becomes  positive immediately after the particle has left the fertile site. To model a refractory period, one could also assume that the fertility rate is zero on an interval containing the fertile site (other models of a refractory period     have been proposed in \cite{refractory}). The second derivative of the stationary density profile at the origin is positive (the probability density of a single RTP was shown to present a minimum at the origin in a transient regime in \cite{malakar2018steady}). 
    The value of this second derivative depends only on the rate of exponential growth of the density of right-movers at the origin. Moreover, it goes to zero quadratically with the fertility rate $K$, for low values of $K$, with a prefactor that depends on the choice of the fertility function (through a single value of its Laplace transform). {{A Gamma-distributed fertility function can be described as resulting from the interaction of RTPs with a bath of active particles 
     that are attracted to the fertile site by a harmonic potential,  subjected to a strong friction, and allowed to jump farther from the fertile site.}}\\

 We took the Laplace transform of the equations of motion w.r.t. the time variable, which calculation yielded the 
  stationary density of particles (normalised by the exponentially-growing number of particles at the origin), without the 
   need for Laplace inversion. This stationary density profile does not depend on the initial conditions.
    From a formal perspective, the rate of exponential growth of the density of right-movers at the origin was 
 obtained from an integral equation, which resembles the renewal equations 
   used to extract the steady state of systems under resetting (see \cite{evans2011diffusion,evans2011optimal,evans2018run,refractory,mercado2018lotka,toledo2019predator,ZRPResetting,IsingResetting,sadekar2020zero,santra2020run,Santra2020,grange2020susceptibility} for examples, as well as \cite{topical} and references therein for a review). The  Laplace transform of the equations of motion was also observed to yield the stationary probability density 
      of a single run-and-tumble particle 
    subjected to resetting in  \cite{evans2018run}.\\

     There is some intuitive analogy between the present model at large times and a system with a fixed number of particles subjected to resetting. Indeed, when the number of particles becomes large in a system with a fertile site, the evolution of the system is  driven by large numbers of 
     newly created particles that are going to flip direction after their creation. The change of direction happens at a characteristic unit distance from the origin,  and 
      directs the particles towards the fertile site. This situation is intuitively equivalent to the resetting of a large fraction of the system to the origin. 
  However,   the steady state we identified at large times is  not the one of the system, as the number of constituents grows indefinitely, but a normalised version, because
     we divided by the exponentially-growing density of particles at the origin. This feature was also observed in \cite{bauer2019random}
      for diffusive particles on a lattice. Taking interactions into account could yield insights on the formation of clusters, as in the crowded  model of swimmers in    \cite{soto2014run}. Intuitively, the clusters could grow from the position of the two peaks observed in the density.\\

  {{In our model, each new particle inherits the velocity of its parent and starts moving immediately after it has been produced. Several 
   modifications of this prescription can be proposed. For example the velocity state of the new particle could be a centered binary variable. This would make the dynamics of each new particle independent of the one of its parent. The creation terms in the equations of motion would be replaced with the average of the two creation terms we wrote in Eq. (\ref{sys}).  Moreover, single-cell observations of swimming and growth properties  of {\emph{E. coli}} \cite{umehara2007origin} revealed that a prolonged pause is taken by cells before  division (the duration of pausing is typically an order of magnitude larger than the duration of tumble events). To model this behaviour, we have to modify the RTP behaviour. At the level of simplification of our model, a  tractable modification consists  in inserting the pause after the cell division. Let us keep the RTP description of parent particles, and let each new particle take a pause before it starts its run-and-tumble motion. The new particles  stay where they have been created (with zero velocity), for some random  time. This random time  is again a refractory period, after which each new particle draws its initial velocity state from a centered binary distribution and starts its RTP dynamics. The probability density $W$ of the refractory period would be a new parameter of the model (the
    previous modification in which the velocity state is drawn at the time of creation corresponds to setting $W$ to a Dirac mass at zero).
      Technically, the corresponding creation terms 
      at time $t$ would correspond to a creation event at time $t-\tau$ (by a particle of positive or negative velocity), followed by a refractory period $\tau$, for some $\tau$ in $[0,t]$.   Following the reasoning of  \cite{refractory}, the creation terms in the equations of motion would be replaced with the convolution of the probability density $W$ and the average of the two creation terms in Eq. (\ref{sys}). As this convolution is in the time variable, the Laplace transform maps the convolution to a product with 
       the Laplace transform $\tilde{W}(s)$. 
       The Laplace transform of the equations of motion would then contain the average of the creation terms in Eq. (\ref{LaplaceSys}), weighted by $\tilde{W}(s)$.
        As the parity of the model is not broken by these modifications (Eq. (\ref{parity}) still holds), every step in the derivation could be followed in a straightforward way.}}\\

       It would be interesting to generalise the model to higher dimensions, as in \cite{santra2020run,Santra2020},
      to see whether the exponential growth of constituents persists. {{However, one-dimensional models of active particles are physically relevant in 
       situations where the motion is confined to narrow channels. Models of active Brownian particles (subjected to both translational and rotational diffusion, as well as to an active force)  have been proposed \cite{locatelli2015active}, yielding estimates of the emptying time of a channel. External biases such as gravity have been considered. One could model an internal bias by setting a fertile site in the channel. Intuitively a strong fertility rate would  block  the channel, but there could be a a critical value of the fertility rate below which emptying times stay finite.}}
       It would also be interesting to see whether a confining potential, or an additional 
       zero-velocity state (as in \cite{basu2020exact}) could qualitatively modify the density profile at large time. {{Moreover, adding interactions and noise to one-dimensional models of RTPs is known to lead to collective phenomena such as the formation of high-density, slow-moving domains \cite{tailleur2008statistical}. It would be interesting to see how these phenomena are changed when a fertile site with small fertility rate is added. Introducing a non-zero death rate of the RTPs could lead to a bounded number of constituents, or to a slower total growth. Collective behaviour 
        of active constituents such as the formation of bacterial colonies \cite{vicsek2001fluctuations,vicsek2012collective} would be biased
       by  the presence of a fertile site modelling a particularly rich spot in the substrate.}}


%
%
%
%
%




{{
\section{Appendix A}
}}

 Let us solve Eq. (\ref{ODE1}) by varying the constant:
 \begin{equation}
  Y(x)=:e^{xM}A(x),
  \end{equation}
   where $A(x)$ is a vector-valued function of $x$. This definition implies
  \begin{equation}
  e^{xM} A'(x) = F(x),
 \end{equation}
  which can be solved if the spectrum of the matrix $M$ is known.\\

The matrix $M $ can be diagonalised as follows:
\begin{equation}
\begin{split}
&M = U^{-1} D U,\\
&D:= \begin{bmatrix} 
  \sqrt{\sigma} & 0 \\
   0  & -\sqrt{\sigma} \\
\end{bmatrix},\;\;\;\;\;\; U:=\frac{1}{2}\sqrt{1+\frac{1}{\sigma}}\begin{bmatrix} 
  \sqrt{\sigma} & 1 \\
   -\sqrt{\sigma} &  1\\
\end{bmatrix}, \;\;\;\;\;\;\;U^{-1} = \frac{1}{\sqrt{1 + \sigma}}\begin{bmatrix} 
  1 & -1 \\
   \sqrt{\sigma}  & \sqrt{\sigma} \\
\end{bmatrix}.
\end{split}
\end{equation}
 The matrix $xM$ is therefore exponentiated as follows:
 \begin{equation}
e^{xM} = U^{-1} e^{xD} U,\;\;\;\;\;\;\;\;\;\;e^{xM} = U^{-1} e^{-xD} U,\;\;\;\;\;\;\;\;\;e^{xD} = \begin{bmatrix} 
  e^{x\sqrt{\sigma}} & 0 \\
   0  & e^{-x\sqrt{\sigma}} \\
\end{bmatrix}.
\end{equation}
Calculating the matrix products yields
\begin{equation}\label{expM}
 e^{xM} = \begin{bmatrix} 
     \cosh( x\sqrt{\sigma} )   &     \frac{1}{\sqrt{\sigma}} \sinh( x\sqrt{\sigma} )     \\
   \sqrt{\sigma}   \sinh( x\sqrt{\sigma} ) &    \cosh( x\sqrt{\sigma} ) \\
\end{bmatrix}.
\end{equation}

 \begin{equation}\label{solution1}
  A(x) = A(0) + \int_0^x  U^{-1} e^{-yD} U F(y) dy,\;\;\;\;\;\;\;Y(x) = e^{xM}A(0) +  U^{-1}\left( \int_{0}^x   e^{(x-y)D} U F(y) dy  \right).
 \end{equation}

Coming back to the original problem of Eq. (\ref{diff2}), we have to fix a vector $A(0)$ with two components.
   We can then extract the first component of the solution  
 from Eq. (\ref{solution1}) to read off $\widetilde{n_+}(x,s)$ in terms of the unknown vector $A(0)$:
 \begin{equation}\label{readOff}
 \begin{split}
  \widetilde{n_\pm}(x,s) &= \left[  e^{xM}A(0) \right][1] + \left[  \int_{0}^x     e^{(x-y)M} G_+ (y,s) dy  \right][ 1 ],\\
  \partial_x\widetilde{n_+}(x,s) &= \left[  e^{xM}A(0) \right][2] + \left[  \int_{0}^x     e^{(x-y)M} G_+ (y,s) dy  \right][ 2 ],\\
  \end{split}
 \end{equation}
 where the arguments in square brackets $[1]$ and $[2]$ respectively denote the first and second components of a vector.
  The vector $G_+$  is defined by substituting the function $g_+$  (defined in Eq. (\ref{defg})) to the function $f$ in the vector $F$ 
   defined in Eq. (\ref{ODE1}):
 \begin{equation}
 G_+ (y,s) :=  \begin{bmatrix} 
  0   \\ 
   g_+(y,s)
\end{bmatrix}.
 \end{equation}
 
  Let us  denote the two components of the vector-valued  integration constant $A(0)$ by $\lambda(s)$ and $\mu(s)$:
  \begin{equation}
  A(0) =: \begin{bmatrix} 
  \lambda(s) \\
  \mu(s)
\end{bmatrix}. 
  \end{equation}
   The relevant matrix product in Eq. (\ref{readOff}) is readily expressed using the exponentiated matrix of Eq. (\ref{expM}).
     It  reads
  \begin{equation}\label{termInit}
     e^{xM}A(0) = \begin{bmatrix} 
    \cosh( x\sqrt{\sigma} )  \lambda(s) +  \frac{1}{\sqrt{\sigma}} \sinh( x\sqrt{\sigma} ) \mu(s)\\  
     \sqrt{\sigma}   \sinh( x\sqrt{\sigma} )   \lambda(s) + \cosh( x\sqrt{\sigma} )  \mu(s)
\end{bmatrix}.
 \end{equation}

  Let us  fix the constants $\lambda(s)$ and $\mu(s)$ by imposing the limit  the Laplace transform of the density of right movers at both spatial infinities:
  \begin{equation}\label{lims}
  \underset{|X|\to \infty}{\lim}\widetilde{n_+}(X,s) = 0.
  \end{equation}

 Consider $X>0$. There are terms  in Eq. (\ref{termInit}) that grow exponentially with $X$:
 \begin{equation}
  \left[  e^{XM}A(0) \right](1)   \underset{X\to+\infty}{\sim} \left(\lambda(s) + \frac{\mu(s)}{\sqrt{\sigma}}\right)  \frac{e^{X\sqrt{\sigma}}}{2},\;\;\;\;\;\;\;\;\;
  \left[  e^{XM}A(0) \right](1) \underset{X\to-\infty}{\sim} \left(\lambda(s) - \frac{\mu(s)}{\sqrt{\sigma}}\right)  \frac{e^{-X\sqrt{\sigma}}}{2}.
 \end{equation}
  
   We have to extract the analogous terms 
  from the integral term in Eq. (\ref{solution1}):
 \begin{equation}
 \int_{0}^X    e^{(X-y)M} G_+ (y,s) dy  = \int_{0}^X    \begin{bmatrix} 
     \cosh( (X-y)\sqrt{\sigma} )   &     \frac{1}{\sqrt{\sigma}} \sinh( (X-y)\sqrt{\sigma} )     \\
   \sqrt{\sigma}   \sinh( (X-y)\sqrt{\sigma} ) &    \cosh( (X-y)\sqrt{\sigma} ) \\
\end{bmatrix}     \begin{bmatrix} 
  0  \\
  g_+(y,s)
\end{bmatrix}    dy.
\end{equation}
The first component of the above vector grows  exponentially with $X$:
 \begin{equation}\label{eq++}
 \left[ \int_{0}^X    e^{(X-y)M} G_+ (y,s) dy  \right][1] = \frac{1}{\sqrt{\sigma}} \int_{0}^X    \sinh\left( (X-y)\sqrt{\sigma}\right) g_+(y,s) dy\underset{X\to+\infty}{\sim}
 \frac{e^{X\sqrt{\sigma}}}{2\sqrt{\sigma}} I_{++},
\end{equation}
 with the following notation:
 \begin{equation}
 I_{++} := \int_{0}^\infty    \exp(-{y\sqrt{\sigma}}) g_+(y,s) dy.
 \end{equation} 
 

  Consider $X<0$. The same reasoning yields the following equivalent of the integral terms when $|X|$ becomes large
  \begin{equation}
 \int_{0}^X    e^{(X-y)M} G_+ (y) dy  = \int_{0}^X    \begin{bmatrix} 
     \cosh( (X-y)\sqrt{\sigma} )   &     \frac{1}{\sqrt{\sigma}} \sinh( (X-y)\sqrt{\sigma} )     \\
   \sqrt{\sigma}   \sinh( (X-y)\sqrt{\sigma} ) &    \cosh( (X-y)\sqrt{\sigma} ) \\
\end{bmatrix}     \begin{bmatrix} 
  0  \\
  g_+(y,s)
\end{bmatrix}    dy.
\end{equation}
 \begin{equation}\label{eq-+}
 \left[ \int_{0}^X    e^{(X-y)M} G_+ (y,s) dy  \right](1) = \frac{1}{\sqrt{\sigma}} \int_{0}^X    \sinh\left( (X-y)\sqrt{\sigma}\right) g_+(y,s) dy\underset{X\to-\infty}{\sim}
 -\frac{e^{-X\sqrt{\sigma}}}{2\sqrt{\sigma}}   I_{-+},
\end{equation}
 where the coefficient is again expressed in integral form
 \begin{equation}
 I_{-+} := \int_{0}^{-\infty}    \exp(+{y\sqrt{\sigma}}) g_+(y,s) dy = -  \int_{0}^{+\infty}    \exp(-{y\sqrt{\sigma}}) g_+(-y,s) dy .
 \end{equation}

 The limits we imposed in Eq. (\ref{lims}) therefore yield the two equations 
 \begin{equation}
  \begin{split}
  \lambda(s) +  \frac{1}{\sqrt{\sigma}}  \mu(s)  + \frac{I_{++}}{\sqrt{\sigma}} &= 0, \\
   \lambda(s) -  \frac{1}{\sqrt{\sigma}}  \mu(s)  - \frac{I_{-+}}{\sqrt{\sigma}} &= 0,\\
 \end{split}
\end{equation}
 hence 
  \begin{equation}\label{lambdamuDef}
  \begin{split}
  \lambda(s) &=  \frac{1}{2\sqrt{\sigma}}\left(  - I_{++} +  I_{-+} \right)= \frac{1}{2\sqrt{\sigma}}\left[  \int_{0}^\infty    \exp(-{y\sqrt{\sigma}}) \left(-g_+(y) - g_+(-y)\right) dy \right], \\
   \mu(s)  &=  - \frac{1}{2}\left(   I_{++} +  I_{-+} \right)= \frac{1}{2}\left[\int_{0}^\infty    \exp(-{y\sqrt{\sigma}})\left(  -g_+(y)  + g_+(-y) \right)dy \right].\\
 \end{split}
\end{equation}
 The corresponding integrals are worked out in  Appendix B.\\
 
  {{
\section{Appendix B}
}}

Let us work out the integrals that appear in the solution of the equations of motion (Eq. (\ref{affineSys})), where we treated $\tilde{R}(s)$ as a parameter.
 They are affine functions of $\tilde{R}(s)$, because of the structure of the function $g_+$ defined in Eq. (\ref{defg})
\begin{equation}
\begin{split}
g_+(y,s) =&    - \left( \frac{s}{2} + 1\right) \varphi(x) + \frac{1}{2}\varphi'(x)\\
        &
         +{\textcolor{black}{ K  \tilde{R}(s) \left[ -   2(s+1)  e^{-(s+1)x} \Theta(x)
        -   e^{(s+1)x} \Theta(- x)  + e^{-(s+1)x} \Theta'(x) \right]}}.
        \end{split}
\end{equation}


The integral $I_{++}$ . We will only need the coefficient of the unknown parameter $K\tilde{R}(s)$. Denoting  by  $ j_{++}(s)$ the value of $I_{++}$ at zero fertility rate (whose explicit expression we will need to work out the prefactor in Eq. \ref{lambdamuDef}), we obtain
\begin{equation}\label{I++}
\begin{split}
 I_{++} &=   \int_{0}^{+\infty}    \exp(-{y\sqrt{s(s+2)}}) g_+(y,s) dy\\
     &= j_{++}(s) +   {\textcolor{black}{ K  \tilde{R}(s)  \int_{0}^{+\infty}  e^{-{y\sqrt{s(s+2)}}} \left[ -   2(s+1)  e^{-(s+1)y} \Theta(y)
        -   e^{(s+1)y} \Theta(- y)  + e^{-(s+1)y} \Theta'(y)\right] dy}}\\
      &=   j_{++}(s) +   {\textcolor{black}{ K  \tilde{R}(s) \left[  -2( s+1)\tilde{\Theta}(s + 1 +\sqrt{s(s+2)})  + \widetilde{\Theta'}(s + 1+\sqrt{s(s+2)}) \right]}}\\
      &=   j_{++}(s) +   {\textcolor{black}{ K  \tilde{R}(s) \left(  -s -1 + \sqrt{s(s+2)} \right)\tilde{\Theta}(s+ 1 +\sqrt{s(s+2)}) }},
 \end{split}
\end{equation}
 where we used the assumption $\Theta(0)=0$ when working out the Laplace transform of $\Theta'$. The term $j_{++}(s)$ reads
 \begin{equation}
 j_{++}(s) = \int_0^\infty e^{-y\sqrt{s(s+2)}} \left[ -\left( \frac{s}{2} +1\right) \varphi(y ) + \frac{1}{2} \varphi'(y ) \right]dy.
 \end{equation}

 Similarly, denoting   by  $ j_{-+}(s)$ the value of $I_{-+}$ at zero fertility rate, we obtain
\begin{equation}\label{I-+}
\begin{split}
 I_{-+} &= \int_{0}^{-\infty}    \exp(+{y\sqrt{s(s+2)}}) g_+(y,s) dy = -  \int_{0}^{+\infty}    \exp(-{y\sqrt{s(s+2)}}) g_+(-y,s) dy \\
   &=   j_{-+}(s) -   {\textcolor{black}{ K  \tilde{R}(s)  \int_{0}^{+\infty}  e^{-{y\sqrt{s(s+2)}}} \left[ -   2(s+1)  e^{(s+1)y} \Theta(-y)
        -   e^{-(s+1)y} \Theta( y)  +  e^{(s+1)y} \Theta'(-y)  \right] dy}}\\
    & =      j_{-+}(s)   {\textcolor{black}{ +K  \tilde{R}(s) \tilde{\Theta}(s+ 1 +\sqrt{s(s+2)})      }},
\end{split} 
\end{equation}
\begin{equation}
\begin{split}
 j_{-+}(s) =& \int_0^{-\infty} e^{y\sqrt{s(s+2)}} \left[ -\left( \frac{s}{2} +1\right) \varphi(y ) + \frac{1}{2} \varphi'(y ) \right]dy\\
 =& -\int_0^{\infty} e^{-y\sqrt{s(s+2)}} \left[ -\left( \frac{s}{2} +1\right) \varphi(-y ) + \frac{1}{2} \varphi'(-y ) \right]dy\\
 =& -\int_0^{\infty} e^{-y\sqrt{s(s+2)}} \left[ -\left( \frac{s}{2} +1\right) \varphi(y ) - \frac{1}{2} \varphi'(-y ) \right]dy,
 \end{split}
 \end{equation}
  where we used the parity of the function $\varphi$ (which implies that $\varphi'$ is odd).\\

From the definitions in Eq. (\ref{lambdamuDef}) we therefore obtain
\begin{equation}\label{lambdaVal}
\begin{split}
\lambda(s) 
  & = \frac{1}{2\sqrt{s(s+2)}}\left(  -j_{++}(s)   +  j_{-+}(s) \right) 
   + {\textcolor{black}{ \frac{ K  \tilde{R}(s)}{2\sqrt{s(s+2)}}\left(  s + 2 - \sqrt{s(s+2)} \right) \tilde{\Theta}(s+ 1 +\sqrt{s(s+2)} ) }},\\
\end{split}
\end{equation}
 The function $\tilde{\psi}(s)$ and $\tilde{\Xi}(s)$ introduced in Eq. (\ref{LaplacePack}) can be read off as reported in Eqs  
  (\ref{psiTilde})  and (\ref{exprXi}). Indeed
\begin{equation}
\tilde{\psi}(s) = \frac{1}{2\sqrt{s(s+2)}}(-j_{++}(s)+ j_{-+}(s))= \frac{1}{2\sqrt{s(s+2)}}\int_0^\infty e^{-y\sqrt{s(s+2)}} (s+2)\varphi(y) dy.
\end{equation}

Moreover,
\begin{equation}\label{muApp}
\begin{split}
\mu(s) & =  - \frac{1}{2}\left(   I_{++} +  I_{-+} \right)\\
 & = -\frac{1}{2}\left(  j_{++}(s)   +  j_{-+}(s) \right) 
   + {\textcolor{black}{\frac{ K  \tilde{R}(s)}{2}  (s- \sqrt{s(s+2)}) \tilde{\Theta}(s+ 1 +\sqrt{s(s+2)} )   }}.
\end{split}
\end{equation}

 The integral term in the expression of the density of right-movers in Eq. (\ref{affineSys}) reads as follows (in the notations of Eq. \ref{notationsJ} we denote by $k(s)$ the value of the 
  integral at zero fertility rate, and we are interested in the coefficient of $\tilde{R}(s)$, denoted by $J(\chi,x)$):
\begin{equation}\label{JApp}
\begin{split}
\mathcal{I}(x,s) =&  \int_{0}^x \sinh\left( (x-y)\sqrt{\sigma}\right) g_+(y) dy\\
=& k(s) \\
 & {\textcolor{black}{  +K  \tilde{R}(s)    \frac{e^{x\sqrt{\sigma}}}{2}\int_0^x   e^{-{y\sqrt{s(s+2)}}}
   \left[ -   2(s+1)  e^{-(s+1)y} \Theta(y)
        -   e^{(s+1)y} \Theta(- y)  +  e^{-(s+1)y} \Theta'(y) 
        \right] dy
   }}\\
 &  {\textcolor{black}{ - K  \tilde{R}(s)   \frac{e^{-x\sqrt{\sigma}}}{2}\int_0^x  e^{+{y\sqrt{s(s+2)}}} \left[ -   2(s+1)  e^{-(s+1)y} \Theta(y)
        -   e^{(s+1)y} \Theta(- y)  +  e^{-(s+1)y} \Theta'(y)  \right] dy }}\\
        =& k(s)   \\
      &  +K  \tilde{R}(s)   \theta( x) ( \frac{e^{x\sqrt{\sigma}}}{2}  \int_0^x   e^{-{y\sqrt{s(s+2)}}}
   \left[ -   2(s+1)  e^{-(s+1)y} \Theta(y)
         +  e^{-(s+1)y} \Theta'(y) 
        \right]   dy   \\
    &    -  \frac{e^{-x\sqrt{\sigma}}}{2}
          \int_0^x  e^{+{y\sqrt{s(s+2)}}} \left[ -   2(s+1)  e^{-(s+1)y} \Theta(y)
       +  e^{-(s+1)y} \Theta'(y)  \right] dy 
         )\\
      & + K  \tilde{R}(s)  \theta( -x)  (   \frac{e^{x\sqrt{\sigma}}}{2} 
       \int_0^x   e^{-{y\sqrt{s(s+2)}}}
   \left[
        -   e^{(s+1)y} \Theta(- y)  
        \right]   dy   \\
  &   - \frac{e^{-x\sqrt{\sigma}}}{2} 
      \int_0^x  e^{+{y\sqrt{s(s+2)}}} \left[ 
       -   e^{(s+1)y} \Theta(- y)     dy 
     \right] ), \\
 \end{split}
 \end{equation}
  where $\theta$,  the Heaviside step function, has been used to deal separately with the case of positive and negative $x$ (using the fact that value of the fertility function $\Theta(x)$ is zero for negative  $x$).

\begin{equation}\label{JApp}
\begin{split}
\mathcal{I}(x,s) =& k(s)   \\
      &  +K  \tilde{R}(s)   \theta( x) ( \frac{e^{x\sqrt{\sigma}}}{2} \left[ e^{-x(\sqrt{\sigma} + s + 1)} \Theta(x) + ( -s-1+\sqrt{s(s+2)}) \int_0^x   e^{-{y( s + 1 +\sqrt{s(s+2)})}} \Theta(y) dy\right]
   \\
    &    -  \frac{e^{-x\sqrt{\sigma}}}{2}
     \left[ e^{+x(\sqrt{\sigma} - s - 1)} \Theta(x) + ( -s-1-\sqrt{s(s+2)}) \int_0^x   e^{-y( s + 1 -\sqrt{s(s+2)})} \Theta(y) dy\right]
         )\\
     & + K  \tilde{R}(s)  \theta( -x)  (  +\frac{e^{x\sqrt{\sigma}}}{2} 
       \int_0^{-x}   e^{ -y(s+1 -\sqrt{s(s+2)})}
    \Theta( y)     dy   \\
  &   - \frac{e^{-x\sqrt{\sigma}}}{2} 
      \int_0^{-x}  e^{-y( s + 1 +\sqrt{s(s+2)})} \Theta(y)     dy ) \\
       =& k(s)   \\
      &  +K  \tilde{R}(s)   \theta( x) (  \frac{e^{-(s+1)x}}{2} \Theta(x) +
      ( -s-1+\sqrt{s(s+2)}) \frac{e^{x\sqrt{\sigma}}}{2}\int_0^x   e^{-{y( s + 1 +\sqrt{s(s+2)})}} \Theta(y) dy
   \\
    & - \frac{e^{-(s+1)x}}{2} \Theta(x) + ( s+1+\sqrt{s(s+2)})  \frac{e^{-x\sqrt{\sigma}}}{2} \int_0^x   e^{-y( s + 1 -\sqrt{s(s+2)})} \Theta(y) dy
         )\\
     & + K  \tilde{R}(s)  \theta( -x)  (  +\frac{e^{x\sqrt{\sigma}}}{2} 
       \int_0^{-x}   e^{ -y(s+1 -\sqrt{s(s+2)})}
    \Theta( y)     dy   \\
  &   - \frac{e^{-x\sqrt{\sigma}}}{2} 
      \int_0^{-x}  e^{-y( s + 1 +\sqrt{s(s+2)})} \Theta(y)     dy ) \\
\end{split}
\end{equation}

Taking the limit of large and positive $x$, we notice that the coefficient of $K  \tilde{R}(s)$ in $\mathcal{I}(x,\sigma)$ is equivalent to
$  ( -s-1+\sqrt{s(s+2)}) \frac{e^{x\sqrt{\sigma}}}{2}\tilde{\Theta}(  s + 1 + \sqrt{s(s+2)})$,
 which is consistent with  the expression of 
 the coefficient of $K  \tilde{R}(s)$ in the expression of $I_{++}$ in Eq. (\ref{I++}), and the equivalent displayed in Eq. (\ref{eq++}).
  Similarly, for large and negative $x$, the coefficient of $K  \tilde{R}(s)$ in $\mathcal{I}(x,\sigma)$ is equivalent to
$ - \frac{e^{-x\sqrt{\sigma}}}{2}\tilde{\Theta}(  s + 1 + \sqrt{s(s+2)})$,
 which is consistent with  the equivalent displayed in Eq. (\ref{eq++}).

%

  {{
    \section{Appendix C}
    }}

      It will be convenient to introduce 
      \begin{equation}\label{DDef}
       D( s,x) := \int_x^{\infty} e^{-sy} \Theta( y ) dy,
      \end{equation}
       in order to make use the properties of the growth rate (Eq. \ref{defChi}). Indeed with this notation
      \begin{equation}       
      C_s(x) = \int_0^x   e^{-sy} \Theta(y) dy = \tilde{\Theta}( s ) - D(a,x).
      \end{equation}
       We will again use the notation $\sigma$, but for the particular value
       \begin{equation}
        s:= \chi, \;\;\;\;\;\; \sigma = \sqrt{s(s+2)}.
       \end{equation}

     \subsection*{Large and positive $x$} Consider $x\gg \sigma^{-1}$. Because of the sign constraint, Eq. (\ref{stationaryxz}) becomes
     \begin{equation}\label{stationaryxzPos}
     \begin{split}
     \tilde{M}(x,\chi) = & \cosh\left(  \sigma x  \right) + \frac{K}{2\sigma}\sinh\left(  \sigma x  \right)(\chi - \sigma) \tilde{\Theta}( \chi + 1 + \sigma) \\
      &+ \frac{K}{2\sigma}( -\chi - 1 + \sigma) e^{\sigma x}\left( \tilde{\Theta}( \chi + 1 + \sigma)  - D( \chi + 1 +\sigma,x)\right) \\
      & + \frac{K}{2\sigma}( \chi + 1 + \sigma)  e^{-\sigma x}\left( \tilde{\Theta}( \chi + 1 - \sigma)  - D( \chi + 1 -\sigma,x)\right) \\
     \end{split}
     \end{equation}
     
     The dominant term in the expression  of $\tilde{M}(x,\chi)$ in Eq. (\ref{stationaryxzPos})
      is proportional to $e^{\sigma x}$, but the coefficient should vanish. It is followed by terms proportional to $e^{-\sigma x}$,
        and terms proportional to $e^{\sigma x} D( \chi+1 + \sigma, x)$, and subdominant terms.\\

        Let us check that the coefficient of $e^{\sigma x}$ vanishes.  It reads 
     \begin{equation}\label{checkPos}
     \begin{split}
     \frac{1}{2}+&\frac{K}{4\sigma}( \chi - \sigma) \tilde{\Theta}( \chi + 1 - \sigma)+ \frac{K}{2\sigma}(-\chi - 1 + \sigma)\tilde{\Theta}( \chi + 1 - \sigma)\\
     &= \frac{1}{2}\left[ 1 + \frac{K}{2}\frac{-\chi - 2 + \sigma}{\sigma}  \tilde{\Theta}( \chi + 1 - \sigma) \right] \\
      & =   \frac{1}{2}\left[ 1 + \frac{K}{2}\left(  1 - \sqrt{\frac{\chi + 2}{\chi}}   \right) \tilde{\Theta}( \chi + 1 - \sigma)\right],     
     \end{split}
     \end{equation}
     which is zero thanks to Eq. (\ref{eqRate}). \\

      Moreover, for a Gamma-distributed fertility function, the function we denoted by $D$ is expressed in terms of the upper incomplete Gamma function:
      \begin{equation}
       D(s,x) = \frac{1}{\Gamma( k ) ( sa + 1)^k} \int_{x(s + a^{-1})}^\infty z^{k-1}e^{-z} dz.
      \end{equation}
      The asymptotic behaviour of the  upper incomplete Gamma function
      \begin{equation}
      \Gamma(k,x) := \int_x^\infty y^{k-1} e^{-y} dy \underset{x\to +\infty}{\sim} x^{k-1} e^{-x}
       \end{equation}
      induces the following equivalent:
      \begin{equation}\label{equivD}
       D( s,x)  \underset{x\to +\infty}{\sim} \frac{1}{\Gamma(k) a^k x( s + a^{-1})}e^{-x(s + a^{-1})}.
      \end{equation}
      
     We can therefore compare  $e^{\sigma x} D( \chi+1 + \sigma, x)$ to $e^{-\sigma x}$ through
     \begin{equation}
      e^{\sigma x} D( \chi+1 + \sigma, x) \underset{x\to +\infty}{\sim} \frac{1}{\Gamma(k) a^k x( \chi+1 + \sigma + a^{-1})}e^{-x(\chi + 1 + a^{-1})}.
     \end{equation}
     Starting from 
     \begin{equation}
      \chi(\chi + 2) < (\chi + 1 )^2, \;\;\;\;\;{\mathrm{hence}} \;\;\;\sigma<\chi + 1,
     \end{equation}
     we obtain 
       \begin{equation}
     \sigma < \chi + 1 + a^{-1},\;\;\;\;\;\;{\mathrm{hence}}\;\;\;\; e^{\sigma x} D( \chi+1 + \sigma, x)\underset{x\to +\infty}{\ll} e^{-\sigma x}.
       \end{equation}
      
      The expression  of Eq. (\ref{stationaryxzPosEq}) therefore decays exponentially at large and positive $x$
       \begin{equation}\label{stationaryxzPosEq}
     \tilde{M}(x,\chi)  \underset{x\to +\infty}{\sim} \left[ \frac{1}{2} + \frac{K}{2\sigma}\left( -\frac{1}{2} (\chi - \sigma) \tilde{\Theta}( \chi + 1 + \sigma)
       + (\chi + 1 + \sigma)\tilde{\Theta}( \chi + 1 - \sigma)\right)   \right]  e^{-\sqrt{\chi( \chi + 2 )}x}.
     \end{equation}

      \subsection*{Large and negative $x$}
      
       Consider a large and negative value of $x$, i.e. $|x|\gg \sigma^{-1}$. Because of the sign constraint, Eq. (\ref{stationaryxz}) becomes
     \begin{equation}\label{stationaryxzNeg}
     \begin{split}
     \tilde{M}(x,\chi) = & \cosh\left(  \sigma x  \right) + \frac{K}{2\sigma}\sinh\left(  \sigma x  \right)(\chi - \sigma) \tilde{\Theta}( \chi + 1 + \sigma) \\
      &+\frac{K}{2\sqrt{\chi( \chi + 2)}}  [  e^{x\sqrt{\chi( \chi + 2 )}}
       \int_0^{-x}   e^{ -y(\chi+1 -\sqrt{\chi (\chi +2)})}
    \Theta( y)     dy   \\
  &   - e^{-x\sqrt{\chi( \chi + 2 )}}
      \int_0^{-x}  e^{-y( \chi  + 1 +\sqrt{\chi (\chi +2)})} \Theta(y)     dy ]\\
      =& \cosh\left(  \sigma x  \right) + \frac{K}{2\sigma}\sinh\left(  \sigma x  \right)(\chi - \sigma) \tilde{\Theta}( \chi + 1 + \sigma) \\
      &+\frac{K}{2\sqrt{\chi( \chi + 2)}}  [  e^{\sigma x} \tilde{\Theta}( \chi + 1 - \sigma) - e^{\sigma x} D( \chi+1 - \sigma, -x))\\
        &-  e^{-\sigma x}\tilde{\Theta}( \chi + 1 + \sigma) + e^{-\sigma x}D( \chi+1 + \sigma, -x)].
     \end{split}
     \end{equation}
      The dominant term in the expression  of 
      $\tilde{M}(x,\chi)$ in Eq. (\ref{stationaryxzPos})
      is proportional to $e^{-\sigma x}$, but the coefficient should vanish. It is followed by terms proportional to $e^{+\sigma x}$,
        and terms proportional to $e^{-\sigma x} D( \chi+1 + \sigma, -x)$, and subdominant terms.\\
      
      Using the equivalent of $D$ in Eq. (\ref{equivD}) yields
      \begin{equation}
      e^{-\sigma x} D( \chi+1 - \sigma, -x)   \underset{x\to -\infty}{\sim} -\frac{1}{\Gamma(k) a^k x( \chi+1 - \sigma + a^{-1})}e^{+x(\chi + 1 +a^{-1})}
      \end{equation}
      Using again $\sigma < \chi + 1 + a^{-1}$ yields 
        \begin{equation}
       e^{-\sigma x} D( \chi+1 - \sigma, -x) \underset{x\to +\infty}{\ll} e^{+\sigma x}
        \end{equation}
        The expression  of Eq. (\ref{stationaryxzPosEq}) therefore decays exponentially at large and negative $x$:
        \begin{equation}\label{stationaryxzPosEq}
     \tilde{M}(x,\chi)  \underset{x\to +\infty}{\sim} \left[ \frac{1}{2} +\frac{K}{2\sigma}\left( \frac{1}{2} (\chi - \sigma) \tilde{\Theta}( \chi + 1 + \sigma)
       + \tilde{\Theta}( \chi + 1 - \sigma)\right)   \right]  e^{+\sqrt{\chi( \chi + 2 )}x}.
     \end{equation}

     Let us check that the coefficient of $e^{-\sigma x}$ vanishes. From Eq. (\ref{stationaryxzNeg}) t reads\\
      \begin{equation}
      \begin{split}
       \frac{1}{2}& - \frac{K}{4\sigma}(\chi - \sigma ) \tilde{\Theta}( \chi + 1 + \sigma) - \frac{K}{2\sigma}\tilde{\Theta}( \chi + 1 + \sigma)\\
       =& \frac{1}{2} \left[ 1 +\frac{K}{2\sigma}  (-\chi + \sigma -2)  \tilde{\Theta}( \chi + 1 + \sigma)   \right]\\
       = & \frac{1}{2} \left[ 1 +\frac{K}{2}\left( 1 - \frac{\chi + 2}{\sigma}\right) \tilde{\Theta}( \chi + 1 + \sigma) \right],
      \end{split}
      \end{equation}
      which matches the coefficient found in Eq. \ref{checkPos} and therefore equals zero.\\

\bibliography{bibRefsNew} 
\bibliographystyle{ieeetr}
         
 \end{document}